\documentclass[twocolumn,showpacs,preprintnumbers,
amsmath,amssymb]{revtex4}
\usepackage{graphicx}

\begin{document}
\title{DILUTE METALS}
\author{V. N. Bogomolov}
\affiliation{A. F. Ioffe Physical \& Technical Institute,\\
Russian Academy of Science,\\
194021 St. Petersburg, Russia} \email{V.Bogomolov@mail.ioffe.ru}
\date{\today}
\begin{abstract}
 Superconductivity $(T_{c}),$ like any other property of a condensate, depends critically on the
concentration of atoms. "Physical" dilution of metals exists in
nonstoichiometric compounds. In such stoichiometric compounds as
oxides, oxygen initiates "chemical" dilution of the metal, but its
efficiency can be estimated only if the real radius of the oxygen
ion, $r_{0} \sim 0.5 \AA,$ is used in the calculation. The
ground-state radii of metal atoms $r_{m} \sim (1.3-2.0 \AA) \gg
r_{0},$ so that atoms of metals occupy in the lattice $\sim 90\%$
of the total volume. Therefore, the lattice parameter and the
electronic properties are determined by the metal-atom ground
states. \linebreak (For $\rm{TiO}_{2},$ the parameter $c = 2.95
\AA \sim 2r_{\textrm{\tiny{Ti}}} = 2.94 \AA;$ for TiO, $a\surd 2 =
5.99 \AA \sim 4r_{\textrm{Ti}} = 5.90 \AA; \linebreak \rm {for
\quad BaTiO}_{3}, \quad c= 4.05 \AA \sim 2r_{\textrm{\tiny{Ba}}} =
4.12 \AA;$ for Y-Ba-Cu-O, $c = 11.68 \AA \sim
(4r_{\textrm{\tiny{Ba}}} + 2r_{\textrm{\tiny{Y}}}) = 11.63 \AA,$
and so on.) Each atomic quantum state can be identified with a
specific physical property of the condensate. As a result of
superposition of the ground and ionic states of each metal atom,
the fraction of the ground states decreases at the expense of the
oxygen-excited ionic ones (thus reducing the effective
concentration of atoms). The bands narrow down (in the limit, to
the electron localization length), with the metal becoming an
insulator, as is the case, for instance, with
$\textrm{YBa}_{2}\textrm{Cu}_{3}\textrm{O}_{8-x}$ as \textit{x}
decreases from $x_{\textrm{\tiny{M}}}$ to $0.$ Conceivably,
reducing the lattice deficiency in oxygen by saturating part of
its valence bonds with H, Li, or B atoms could increase the $T_{c}
\enskip (\textrm{YBa}_{2}\textrm{Cu}_{3}\textrm{O}_{7-x}  \sim
\textrm{YBa}_{2}\textrm{Cu}_{3}\textrm{O}_{7-x}(1
+x)\textrm{H}_{2}\textrm{O}).$ "Self-dilution" of lattices of
excited atoms by ground-state atoms was observed to occur in Pd
and inert gases.
\end{abstract}
\maketitle
\bigskip
   \section{Introduction}

   In describing the properties of oxides, one traditionally assumes
   the radius of the $\rm{O}^{2-}$ ion to be $1.4 \AA,$
although about six different figures, ranging from $1.36\AA$ to
$1.76 \AA,$ exists for this quantity. One should therefore treat
atoms of metals in oxides as ions with $r_{im} \sim (0.7-0.3)
\AA,$ because the interatomic separations constitute ordinarily
$(1.7-2.2) \AA.$ On the other hand, both quantummechanical
calculations \cite{bib1} and experimental values of the
$\rm{O}^{2-}$ radius (in the $\rm{O}_{2}$ molecule) yield $\sim
0.5 \AA$ (the $2p^{6}$ state). When using $0.5 \AA$ for the radius
of the $\rm{O}^{2-}$\!ion, the sublattices of the metal atoms will
be governed by their ground-state radii, which likewise are close
to both calculated \cite{bib1} and experimental values. The
ground-state radii of metal atoms, $r_{m} \sim (1.3-2.0) \AA \gg
r_{0},$ the radii of both the atomic and ionic states of the
oxygen atom $(0.50 \AA),$ which occupies $\sim5\%-7\%$ of the
lattice volume. Therefore, the lattice parameters and electronic
properties are dominated by sublattices made up of ground-state
metal atoms. \linebreak (For $\rm{TiO}_{2},$ the parameter $c =
2.95 \AA \sim 2r_{\textrm{\tiny{Ti}}} = 2.94 \AA;$ \linebreak  for
TiO,\qquad$a\surd 2 = 5.99 \AA \sim 4r_{\textrm{\tiny{Ti}}} = 5.90
\AA; \linebreak \rm{for \quad BaTiO}_{3},\quad c= 4.05 \AA \sim
2r_{\textrm{\tiny{Ba}}} = 4.12 \AA;$ \linebreak for Y-Ba-Cu-O, $c
= 11.68 \AA \sim (4r_{\textrm{\tiny{Ba}}} +
2r_{\textrm{\tiny{Y}}}) = 11.63 \AA,$ and so on.)

    Adhering to the traditional version (1.4 \AA) has not until recently been critical due to a fortuitous matching
of the radii$(r_{\textrm{O}^{2-}} \sim r_{im}).$ In the case of
superconductivity, however, the version of \!$1.4\AA$ \!differs
radically from that of \!$0.5 \AA.$

    A systematic investigation of the properties of condensates in relation to the distances (interaction) among
atoms started apparently sometime after\!$1946,$\!when it was
conjectured that dilution of a metal should increase its
$T_{c}$\!\cite{bib2}. In the $1980$\!s, a series of studies of
metals diluted by atoms of inert gases (IG) were carried out in
connection with the Mott transition (see, e.g., \cite{bib3,bib4}).
These investigations did not, however, yield unambiguous results,
because "mechanical" dilution of a metal (by ammonia \cite{bib2}
or IG atoms) precipitated it rapidly into a separate phase. As
atoms with valence electron pairs are brought close together, the
binding energy of the electrons in a pair decreases gradually, and
their overlap increases. The possibility of the onset of
superconductivity involving localized electronic states in the
vicinity of the Mott transition (gas of atoms-the Mott
insulator-metal) has long been a subject of debate in its various
aspects (see, e.g., \cite{bib5}). To approach this issue
experimentally, one has first to unravel the mechanisms governing
the variation of the effective atom concentration in a substance
and to learn how to control them properly.

   "Physical" (spatial, direct) dilution of metals exists in nonstoichiometric interstitial compounds
(nanocomposites, in which, however, metal-insulator-type contact
interaction between components exists too).

   In oxide-type stoichiometric compounds, "chemical" (indirect) dilution is realized. Each atomic quantum state,
excited or ionized as well, in a condensate can be identified with
a specific physical property. The lattice structure and parameter
and the binding energy are determined by the ground and ionic
quantum states of atoms. The appearance in the superposition of
quantum states of each metal atom of ionic states, which form in
the condensate as its atoms interact with oxygen, brings about a
decrease in the volume density of ground-state metal atoms (i.e.,
in the number of such states per unit volume). The latter is
equivalent to a decrease of the effective atomic concentration
("chemical" dilution). This gives rise to a band narrowing (in the
limit, until complete electron localization obtains), a change in
the electron and phonon properties, and to the metal shifting to
the insulating state. Such evolution of the properties was
observed to occur in
$\textrm{YBa}_{2}\textrm{Cu}_{3}\textrm{O}_{8-x}$ as \it{x} \rm is
reduced from some $x_{\textrm{\tiny{M}}}$ to $0,$ i.e., under
oxidation of a partially reduced metal in an oxide. An increase in
oxygen concentration reduces the volume density of ground states
of metal atoms (the effective concentration of atoms) and changes
the state of the substance (metal-superconductor-insulator). The
potential of increasing $(T_{c})$ in this way is limited by the
rigid constraints of spatial structures and local inhomogeneity of
$\rm{O}^{2-}$ acceptor distribution for $x > 0.$ Uniform increase
of the YBaCuO lattice parameter by replacing the O atoms with the
larger Se atoms enhanced the dilution of the Y, Ba, Cu sublattice
and increased $T_{c}$ to $371\rm{K}$ \cite{bib6}, but at the
expense of making the system unstable. A similar effect was
observed by substituting smaller atoms (Ag) for part of the metal
atoms, although $T_{c}$ rose to $120\rm{K}$ \cite{bib7}.

   "Self-dilution" of the same type was found to occur in Pd and IG condensates. In this case, interaction
creates excited states, and the lattice of atoms in excited states
(i.e., of a larger radius) is diluted by ground-state atoms
\cite{bib8}.

\section{"Physical" dilution (nonstoichiometric interstitial compounds)}
    Metal dilution which does not involve noticeable chemical interactions exists in nonstoichiometric
interstitial compounds (nanocomposites). Consider, for instance,
$\textrm{MeB}_{6}$\!\cite{bib9}. The lattice parameter of the
insulator $\textrm{B}_{6} \,(a = 4.15 \AA)$ varies depending on
the diameter \!$d_{m}$ of the metal atom inserted into the
$\textrm{B}_{6}$ cage. The ratio $a/d_{m}$ for the metals Ce, La,
Ca, Sr, and Ba assumes the following values: $4.15/3.26,
4.15/3.74, 4.15/3.95, 4.19/4.30,$ and $4.27/4.34.$ In \cite{bib9}
were taken for $d_{m}$ the diameters of atoms in the corresponding
metals with a packing density of $0.74.$ The same effect is
observed in $\rm{MeTiO}_{3}$ stoichiometric compounds (see below).
If Ca in the $\rm{CaB}_{6}$ insulator is replaced by atoms of a
trivalent metal, the compounds obtained will be metallic
\cite{bib9}. This justifies the use of metallic diameters. This
situation is shown schematically in Fig.1. The interatomic
distances $d_{m}$ are those of the divalent metal Ca. The packing
density of Ca atoms in the $\rm{B}_{6}$ matrix dropped to $0.52,$
the bands became narrower, and Ca in the simple cubic lattice
became a band insulator with effective interatomic separations
$d^{*}_{\rm{mB}_{6}}.$ Substituting trivalent metals for Ca
results in a half-filled band, with $\rm{MeB}_{6}$ acquiring
metallic properties. A similar situation is observed in
$\rm{MeWO}_{3}$ tungsten bronzes.

   One more aspect of interest is the contact of a lattice of metal atoms with
   an insulator matrix (the case of
metal atoms dispersed in an effective dielectric medium). As is
well known from studies of capillary phenomena \cite{bib10,bib11},
the metal-insulator contact may give rise to a strong interphase
interaction.

   Dilution of Na in $\rm{NH}_{3}$\!\cite{bib2} and in $\rm{WO}_{3} \: (\rm{Na}_{0.05}\rm{WO}_{3})$ \cite{bib12} brought about the onset of
   superconductivity at about 200 and $90\rm{K}$. In both cases, the systems were extremely unstable, because
   the superconductivity set in apparently
when precipitation of the metal phase passed the stage of
formation of $\textrm{Na}_{2}$ molecules with a low concentration
\cite{bib13}.

    \section{"Self-dilution" (monatomic condensates)}
    Interaction of atoms may result in formation not of ionized but of excited quantum states only. This
situation occurs in metallic Pd and in IG condensates
\cite{bib8,bib14}. Atoms of Pd and of the IGs have filled
electronic shells. The Pd ground-state atom $(4d^{10}5s^{0})$ has
a radius $\sim 0.57 \AA$\! \cite{bib1}. The radius of the atom in
the metal, $r_{m} = 1.376 \AA,$ is equal to that of the excited
state atom $(4d^{9}5s^{1}).$ While "chemical" dilution gives rise
to a superposition of the ground and ionized states, here the atom
resides in a superposition of the ground and excited states. The
radii of the states are, however, strongly different in this case
too. The lattice parameter and the binding energy are determined
by the state having the larger radius (the new, excited one). The
old (ground) state of a small radius realizes in the form of
"quantum cavities and channels", which permit high-rate diffusion,
for instance, of hydrogen. The same effect is observed under
"chemical" dilution in oxides, but it involves ionic-type states
$(\rm{TiO}_{2},$\! see below).

   "Self-dilution" occurs in IG condensates too. The state of IG atoms in condensates is also determined by a
superposition of the ground $(r_{g} \sim 0.3-0.6 \AA)$ and excited
$(r_{e} \sim 1.5-2.5 \AA)$ states. The only difference from Pd
consists in the magnitude of the excitation energy. For Pd atoms,
it is
 $\sim 1.3$eV, while for the IG atoms, it is $\sim 10-20$eV. Therefore Pd is a metal in which the
 effective concentration of excited states is $\sim 75\%,$ while that of the ground states ("cavities"),
 only $\sim 25\%.$ For the IG condensates, these states are in the inverse ratio, which makes
 them insulators, but not of the band type \cite{bib14}. For instance, the probability of formation in He
 of excited states $(\rm{He}^{*})$ governing the binding energy and lattice parameter is $\sim 10^{-3}.$
 This probability is certainly not high enough even for crystallization of the condensate.
 Most of the atoms $(99.9\%)$ make up essentially a gas of ground-state noninteracting atoms.
 This figure characterizes the "self-dilution" of He, which is treated by us, however, as a
 liquid of particles with a radius $\sim 1.5\AA,$ although their number is extremely small.
 Compression increases their number, and He undergoes solidification.

   IG condensates are convenient objects for studying the dependence of the properties of a substance on atomic
concentration under compression, including the superconductivity.
In monatomic systems (Pd, IG), excited states (*) form in pair
interactions. "Virtual" diatomic molecules (condensate as a
superposition of the ground and molecular states) are produced.
For instance, $(\textrm{Xe}^{*})_{2}$ is an analog of
$\textrm{Cs}_{2}$ molecules, which have two paired electrons each.
The condensate consists of divalent particles whose concentration
depends on the excitation energy and pressure. Their number in Pd
is large enough to form a metal, while in He it is too small even
to initiate crystallization. Therefore, IG condensates may be
considered as strongly diluted, rarefied gases of divalent metal
atoms (in the state preceding Mott's transition) embedded in a
medium of weakly bound atoms in the ground state (an insulator).
In He these are $(\rm{He}^{4*})_{2}$ in a medium of $\rm{He}^{4}$
atoms, or $(\rm{He}^{3*})_{2}$ in a medium of $\rm{He}^{3}$ atoms.
The pressure-induced conductivity (or superconductivity) appears
initially as a result of formation of a disordered $3D$
percolation "cobweb" of chains of $(\rm{IG}^{*})_{2}$ molecules
\cite{bib8,bib14}.

    \section{"Chemical dilution" (stoichiometric compounds)}
    "Physical" dilution meets with the problem of stabilization of metal atoms present in a low and uniform
concentration. In oxides, oxygen is involved in both the "physical" (spatial) dilution of metal atoms and
"chemical". The latter stems from the fact that the diameter of the ionized states (produced in interaction with
oxygen) taking part in the superposition of quantum states of metal atoms is much smaller than that of the ground
states determining the lattice parameter (volume). This brings about a decrease in the volume density of ground
states (in the limit, to their localization), which is equivalent to a decrease of the effective concentration
of metal atoms (dilution).

   $\textrm{TiO}_{2}$ (rutile) may serve as a good illustration. Figure 2 shows a cut of the $\textrm{TiO}_{2}$
   tetragonal cell in two directions. The radius of the Ti atom $r_{\rm{Ti}} = 1.48\AA,$
   that of the $\textrm{Ti}^{4+}\quad \rm{ion,} r_{i} = 0.48 \AA,$ and of the oxygen ion $\rm{O}^{2-},\quad  r_{O}\sim 0.55\AA $ \cite{bib1}.
   In the metal, $r_{mTi} = 1.46\AA.$ Ti atoms occupy about $95\%$ of atomic volume in the $\rm{TiO}_{2}$ lattice.
   Therefore, both the lattice parameter and the electronic properties of this oxide are governed by the
   sublattice of Ti atoms (in the ground quantum state). The oxygen atoms (acceptors, $\sim 5\%$
   of atomic volume) excite ionic-type states, thus reducing the fraction of ground states in the
   superposition of these Ti atomic states. Interatomic distances in the Ti atom chains (Fig. 2)
   are the same as in the Ti metal (parameter $c = 2r_{Ti} = 2.95\AA).$ The decrease of the ground-state
   concentration in the chains is, however, equivalent to an increase of atomic separations.
   This brings about a narrowing of the electronic bands and formation of an insulator with a
   bandgap $\sim 3\textrm{eV}$ (see Fig. 1). In TiO (Fig. 2), the amount of oxygen is one half that in
   $\rm{TiO}_{2},$
   and the Ti chains are,accordingly, less diluted (TiO is a semimetal, Fig. 1).

   The new, ionic-type quantum states of Ti atoms in $\rm{TiO}_{2},$ not only have the Ti-O binding energy
    but give rise to the formation in Ti chains of "quantum cavities", over which, for instance,
    B or Li can diffuse with a
coefficient of $(10^{-3}-10^{-4})\rm{cm}^{2}/s$
\cite{bib15,bib16}, a figure five to six orders of magnitude
higher than those characteristic of conventional solids. In Pd,
hydrogen diffuses over small-radius ground-state chains among the
excited state atoms \cite{bib8}.

   Application of this concept to $\textrm{YBa}_{2}\textrm{Cu}_{3}\textrm{O}_{8-x}$ suggests the existence of a regular 3D network of "dilute" chains
of Y and Ba atoms \cite{bib17}. This $3D$ conducting network
("gossamer" by \cite{bib18}) is embedded in a medium of Cu atoms
"diluted" by oxygen to the insulator state. As \textit{x} is
varied from $x_{\rm{M}}$ to $0,$ the volume density of
ground-state metal atoms (the effective atom concentration)
decreases, with the compound crossing the
metal-superconductor-insulator stages (Fig. 1). At $x = 0,$ the
degree of Y and Ba dilution is apparently so high as to make the
$\textrm{YBa}_{2}\textrm{Cu}_{3}\textrm{O}_{8}$ lattice close to
unstable.

   Figure 3 shows cuts of the $\rm{CaTiO}_{3}$ and $\rm{BaTiO}_{3}$ unit cells. As in the case of nonstoichiometric
   $\rm{MeB}_{6}$ compounds, the Ba atom stretched the perovskite cell from $c = 3.88 \AA$ in ${\rm{CaTiO}_{3}}$
   to $c = 4.12\AA$ for $\rm{BaTiO}_{3},$ thus providing a
possibility for oxygen ions to displace between the Ti atoms
within $\sim 0.2 \AA.$ The Ba atom chains are "diluted" by oxygen
in the simple cubic cell to the band insulator state (Fig. 1).

 \section{Conclusion}
    Changing the effective concentration of metal atoms is only one of methods to increase the superconducting
transition temperature $T_{c}.$ It is conceivable that a
superconducting state with a high $T_{c}$ can exist in the
vicinity of Mott's transition between the insulator and the metal.
Such successive transitions are known to occur in HTSC materials
at a change of the oxygen concentration in them, i.e., under
variation of the degree of "chemical" dilution of the metal. Such
doping is locally inhomogeneous. However, direct methods capable
of producing stable systems out of uniformly diluted metal atoms
are unknown or still in the stage of development (for instance,
with the use of zeolites as inert diluters \cite{bib19}). Indirect
methods of diluting metals can be properly understood only when
using realistic atomic and ionic radii and after establishment of
a clear correspondence between the physical properties of
condensates and the quantum states of atoms.
   Does development of superconductors with $T_{c} > 300K$ appear feasible? Thus far, reports on such substances
built only on measurements of the conductivity, with which ac
measurements of the magnetic properties may also be classed
($371K$\cite{bib6}). As in \cite{bib2}, the samples were extremely
unstable and poorly reproducible. If, however, the above
explanations of the reasons accounting for the instability are
correct, critical temperatures $T_{c} > 300K$ may be reachable. To
attain uniform "chemical" dilution of metals (as in substitution
of Se for O \cite{bib6}), one would have to use an element with
the ionic radius of oxygen but with a valence of $\sim 2(7/8) =
1.75.$ Another method is that of valence bond saturation. This is,
for instance, the use of such atoms as H, Li, or B as
interstitials. While the system will remain inhomogeneous, the
extent of this inhomogeneity may turn out lower than under oxygen
deficiency (OH or $\rm{H}_{2}\rm{O}$ in place of an empty O site).
In this case, $\rm{YBa}_{2}\rm{H}_{2+x}\rm{Cu}_{3}\rm{O}_{8}$ will
correspond actually to
$\textrm{YBa}_{2}\textrm{Cu}_{3}\textrm{O}_{7-x}
(1+x)(\rm{H}_{2}\rm{O}) \sim
\textrm{YBa}_{2}\textrm{Cu}_{3}\textrm{O}_{7-x}$. An experimental
investigation of the dependence of $T_{c}$ on interatomic
distances (or atomic concentration) would require a quantitative
estimation of the extent of the "dilution" of the metal with
oxygen.
   The concentration of divalent particles in pressure-metallized xenon is stable, uniform, and controllable. No
magnetic measurements have thus far, however, been carried out
\cite{bib14}. It is known that sulfur and oxygen become
superconductors with $T_{c} \: \rm{below} \: \sim 17\rm{K,}$ but
their atoms are paramagnetic.

\begin{figure*}[tp]
\includegraphics[width=\linewidth]{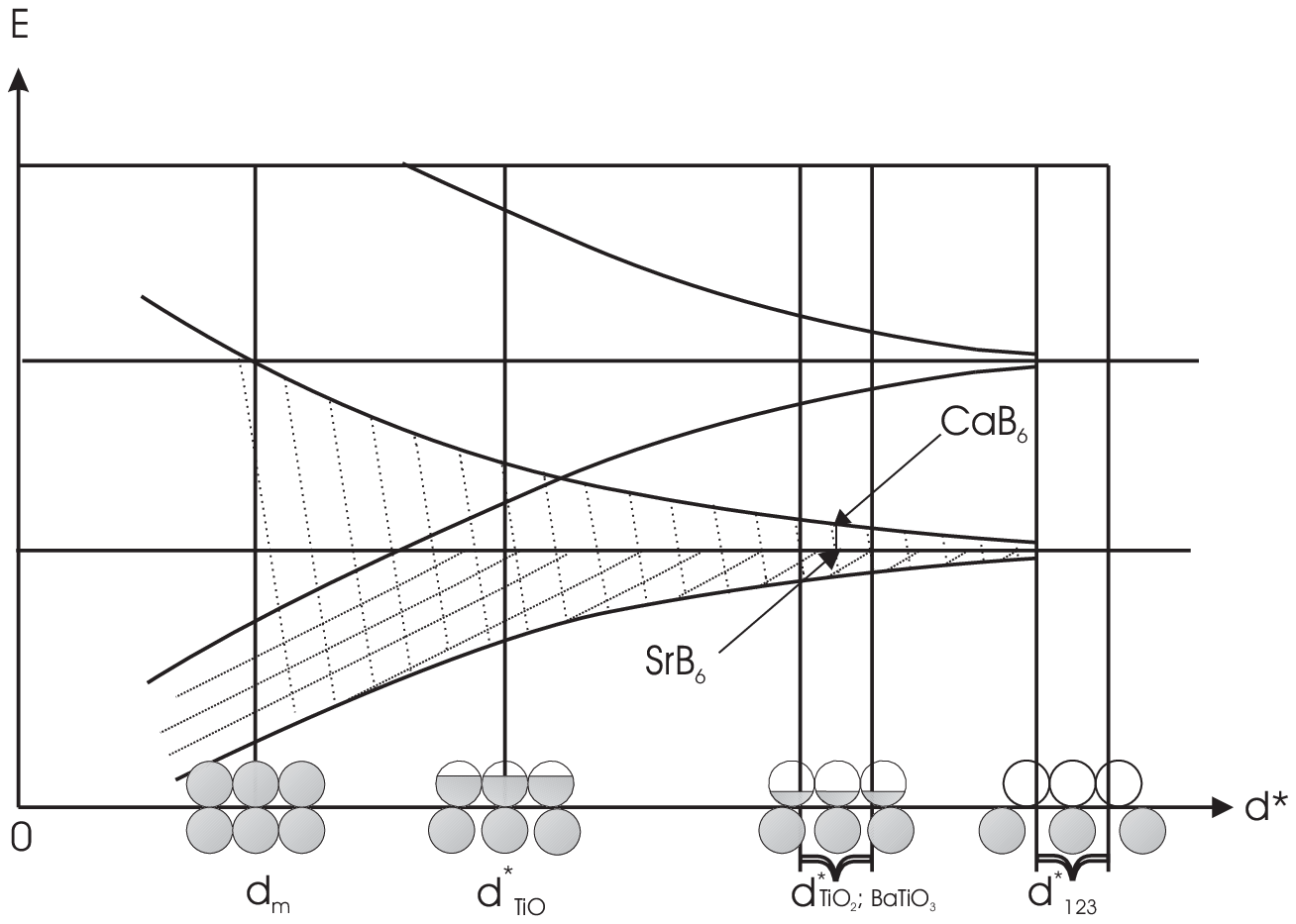}
\caption{Schematic of the conceived dependence of states of a
substance on the effective distances $d^{*}$ between metal atoms
in an oxide. $D^{*}_{123}$ varies from
$d^{*}_{\textrm{YBa}_{2}\textrm{Cu}_{3}\textrm{O}_{7-x}}$ to
$d^{*}_{\textrm{YBa}_{2}\textrm{Cu}_{3}\textrm{O}_{8}}.$}
\label{fig1}
\end{figure*}

\begin{figure*}[p]
\includegraphics[width=\linewidth]{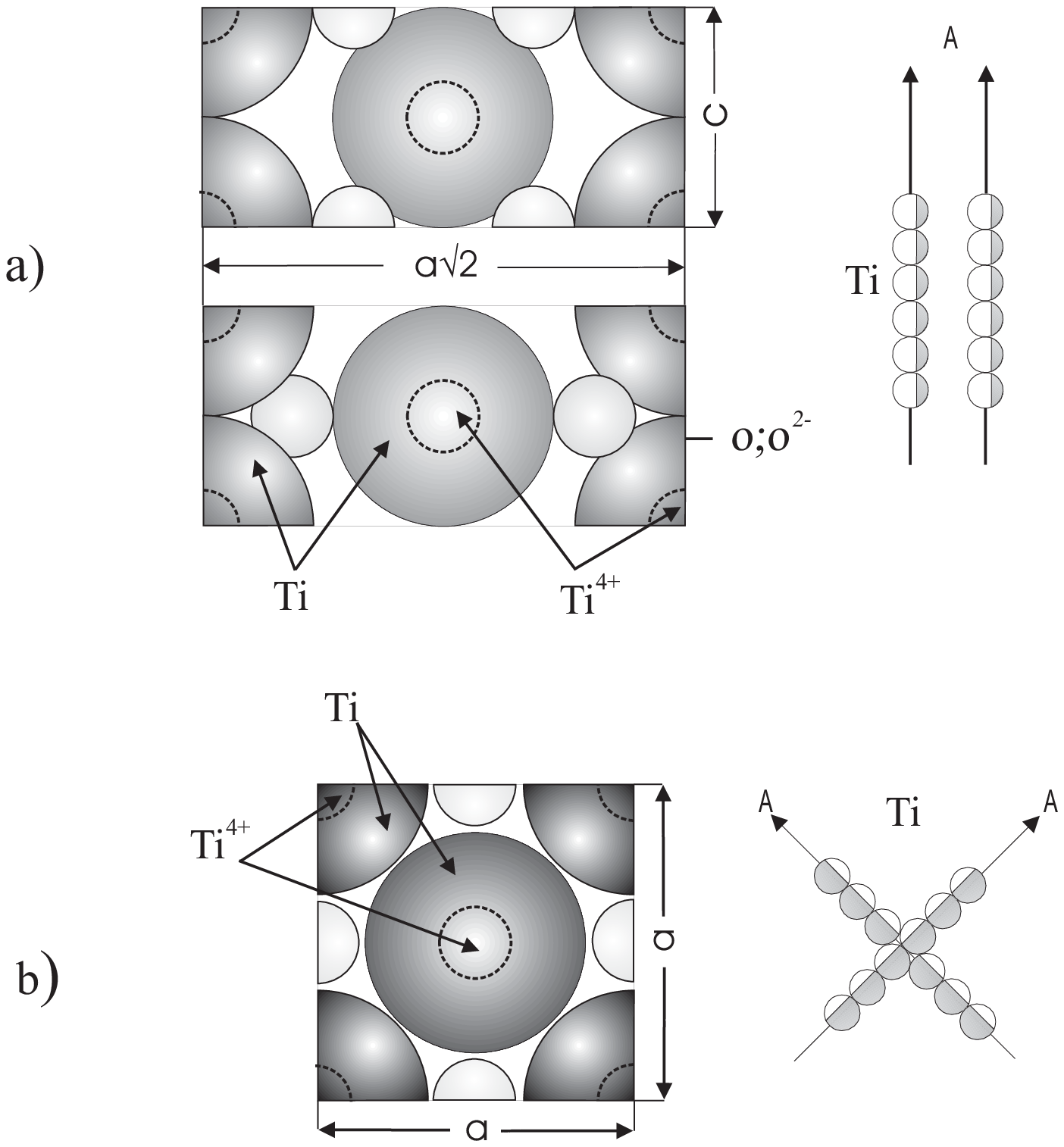}% Here is how to import EPS art
%\caption{\label{fig:wide}
\caption{\\(a). Cut of the $\textrm{TiO}_{2}$ insulator cell in
perpendicular directions; (b) (\textit{a-a}) face of the TiO
semimetal cell. A---direction of the chains of Ti atoms diluted by
oxygen. } \label{fig2}
\end{figure*}

\begin{figure*}[p]
\includegraphics[width=\linewidth]{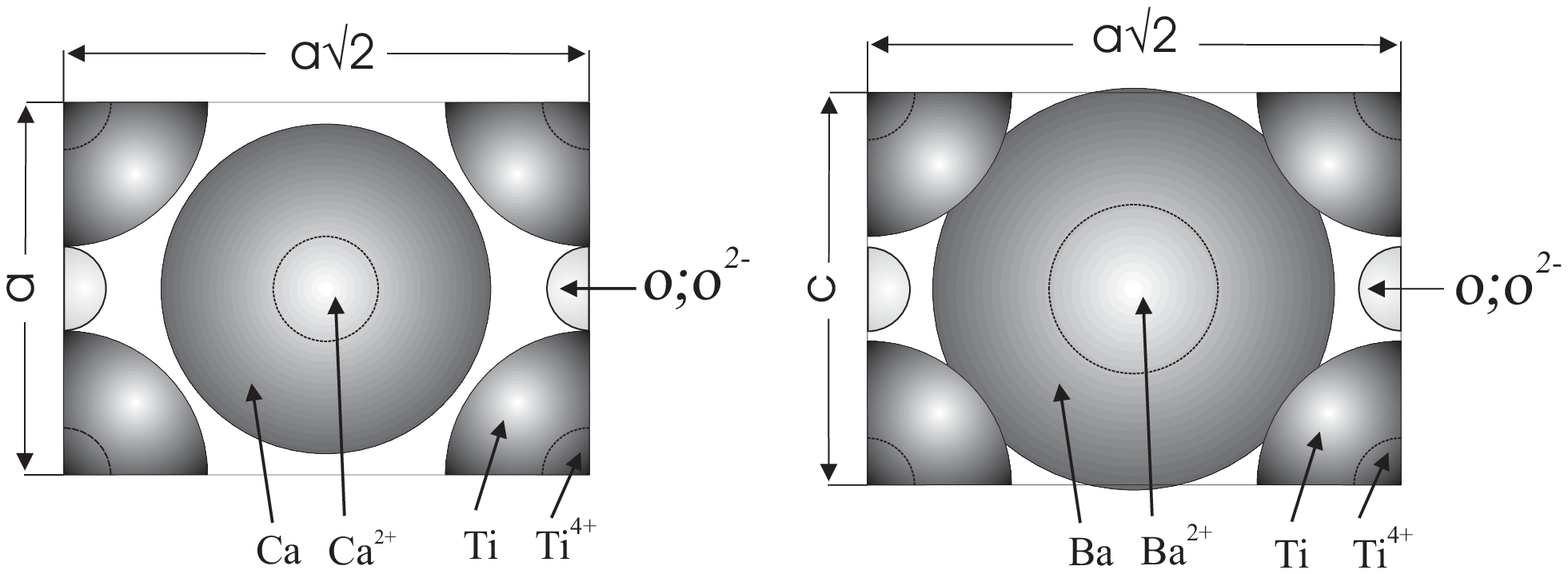}
\caption{Cuts of the $\textrm{CaTiO}_{3}$ and $\textrm{BaTiO}_{3}$
unit cells.}
\label{fig3}
\end{figure*}


\begin{thebibliography}{19}
\bibitem{bib1}   J.T.Waber,Don T. Cromer, J. Chem. Phys.,  {\bf 42},\textit{(12)}, 4116 (1965).
\bibitem{bib2}  R.A.Ogg, Jr. ,    Phys. Rev.   {\bf 69},  \textit{(8)},  243  (1946).
\bibitem{bib3}  R.A.Tilton and C.P.Flynn,   Phys.Rev.Lett., {\bf 34} \textit{( 1 )}, 20  (1975).
\bibitem{bib4} D.J.Phelps, T.Avci and C.P.Flynn, Phys.Rev.Lett., {\bf 34} \textit{(1)}, 23  (1975).
\bibitem{bib5} E.K.Kudinov,    Fizika Tverdogo Tela, {\bf 44}, \textit{(6)}  667 (2002).
\bibitem{bib6} Shabetnik,V.D., Butuzov, S.Yu.,  Plaksii, V.I.,     Techn. Phys. Lett.   {\bf 21} ( 10 ) ,  382   (1995).
\bibitem{bib7}   E.Yanmaz,  I.H.Multu, T.Kucukomeroglu,  M. Altunbas, Supercond. Sci. Technol. {\bf 7}, 903   (1994).
\bibitem{bib8} V.N.Bogomolov,  Techn.Phys.Lett.,  {\bf 28}, \textit{( 3 )} 211 (2002); \textit{Excimer
                interction at the condensation, adsorption and catalysis.}
                Preprint 1536, (Russian Academy of Science, A.F.Ioffe PTI, Leningrad, 1991)
                e-print http://xxx.lanl.gov/abs/cond-mat/9912034; /0302034.
\bibitem{bib9} Mandelcorn,L. (ed) \textit{Non-Stoichiometric Compounds} (Academic Press, N.Y. and London,  1964)
\bibitem{bib10} V.N.Bogomolov,   Phys. Solid State,   {\bf 35} \textit{(4)}, 469 (1993).
\bibitem{bib11} V.N.Bogomolov,   Phys. Rev. B, {\bf 51} \textit{( 23 )}, 17040 (1995).
\bibitem{bib12} A.Shengelaya,  S.Reich,  Y.Tsabba  and K.A. Muller,    Eur. Phys.J.{\bf B12},  13   (1999).
\bibitem{bib13} V.N.Bogomolov, e-print http://xxx.lanl.gov/abs/cond-mat/0304561; 0311265
\bibitem{bib14}  V.N.Bogomolov, \textit{Metallic xenon. Conductivity or Superconductivity?},  Preprint
                1734,(Russian Academy of Science, A.F.Ioffe PTI, St-Petersburg,1999); Techn.Phys.Lett.,
                {\bf 21}, (11) {\bf 928 (1995)}; e-print http://xxx.lanl.gov/abs/cond-mat/9902353;
                 Techn.Phys.Lett.,  {\bf 28} \textit{(3)}, 211 (2002).
\bibitem{bib15} V.N.Bogomolov,     Sov. Phys. Sol. State, {\bf 5} \textit{(7)}, 1468  (1964).
\bibitem{bib16} O.W.Johnson,  Phys. Rev., {\bf 136} \textit{(10)}, 284 (1964).
\bibitem{bib17} V.N.Bogomolov,e-print http://xxx.lanl.gov/abs/cond-mat/0406564.
\bibitem{bib18} R.B.Laughlin, e-print http://xxx.lanl.gov/abs/cond-mat/0209269.
\bibitem{bib19} R. Arita, T. Miyake, T. Kotan,  M.van Schilfgaarde, T. Oka,  K. Kuroki, Y. Nozue, H. Aoki,
                   Electronic properties of alcali-metal loaded zeolites-"supercrystal" Mott insulator.
                   e-print http://xxx.lanl.gov/abs/cond-mat/0304322.
\end{thebibliography}
\end{document}